\documentstyle[aps]{revtex}
\begin{document}
\tightenlines
\title{Inflation and Nonequilibrium Thermodynamics for
the fluctuations in the infrared sector}
\author{Mauricio Bellini\footnote{E-mail address: mbellini@mdp.edu.ar}}
\address{Departamento de F\'{\i}sica, Facultad de Ciencias Exactas  y
Naturales \\ Universidad Nacional de Mar del Plata, \\
Funes 3350, (7600) Mar del Plata, Buenos Aires, Argentina.}
\maketitle
\begin{abstract}
In the framework of inflationary cosmology I study some aspects of
nonequilibrium thermodynamics for the matter field fluctuations.
The thermodynamic analysis is developed for de Sitter
and power - law expansions of the universe. In both cases, I find
that the heat capacity is negative leading respectively, to exponential
and superexponential growth for the number of states in the infrared
sector for de Sitter and power-law expansions of the universe.
The spectrum for the matter field fluctuations can be understood
from the background effective temperature when the horizon entry.
\end{abstract}
\vskip 2cm
\noindent
Pacs number(s): 98.80.Hw, 05.70.Ln\\
\vskip 2cm
\twocolumn
Since inflation stretches microscopic scales into astronomical ones,
it suggests that the density perturbations which provide the seeds
for galaxy formation might have originated as microscopic quantum
fluctuations\cite{lin,star}.

A promising approach towards a better understanding
of these phenomena is the paradigm of stochastic inflation.
The most
widely accepted approach assume that the inflationary phase is driving by a
quantum scalar field $\varphi $ with a potential $V(\varphi )$. Within this
perspective, the stochastic inflation proposes to describe the dynamics of
this quantum field on the basis of a splitting of $\varphi$ in a homogeneous
and an inhomogeneous components. Usually the homogeneous one is interpreted
as a classical field that arises from a coarse-grained average over a
volume larger than the horizon volume, and plays the role of a global
order parameter\cite{goncharov}. All information on scales smaller than this
volume, such as the density fluctuations, is contained in the inhomogeneous
component. During inflation vacuum fluctuations on scales less than
the Hubble radius are magnified into classical perturbations in the
scalar fields on scales larger than the Hubble radius.
The primordial perturbations arise solely
from the zero - point fluctuations of the quantized fields. Although
the region which ultimately expanded to become the observed universe
may have contained excitations above the vacuum, these
excitations would not have any significant effect on the present state
of the universe because a sufficiently large amount of the inflation would
have redshifted
these excitations to immeasurably long wavelengths.
The zero - point fluctuations, on the other hand, have arbitrarily
small wavelengths, indeed they are most significant at very
small length scales. The zero - point fluctuations are imperceptible
because of their short wavelengths, but the
process of inflation can strech these wavelengths to macroscopic and
eventually to astronomical dimensions.
Hence, the density perturbations should be responsible for the
large scale structure formation in the universe.

In this paper I am interested in the study of the termodynamical properties
of matter field fluctuations
in the infrared (IR) sector.
Some thermodynamical researchs were developed recently
in the framework of the density topology of the spacetime
foam\cite{c} and the quantum structure of Schwarschild
black holes\cite{p}.
During inflation this sector varies the
number of degrees of freedom. This is an unstable sector which
describes the universe on a scale much larger than the observable
universe.
A natural
consequence of this approach is the self - reproduction of universes
and the return to a global stationary picture. For inflation the simplest
assumption is that there are two scales: a long - time, long - scale
associated with the vacuum energy dynamics, and the short - time,
short - distance scale associated with a random force component. The
Hubble time $1/H$, separates the two regimes.

The dynamics of a scalar field minimally coupled to a classical gravitational
one is described by the Lagrangian
\begin{equation}
{\cal L}(\varphi,\varphi_{,\mu})=-\sqrt{-g}\left[ \frac{R}{16\pi}+
\frac{1}{2}g^{\mu\nu}\varphi_{,\mu}\varphi_{,\nu} + V(\varphi)\right],
\end{equation}
where $R$ is the scalar curvature $g^{\mu\nu}$ are the components
of the metric tensor (with $\mu,\nu = 0,1,2,3$).

The equation that describes the fluctuations during the inflationary
phase is\cite{Hab,PRD96}
\begin{equation}\label{1}
\ddot\phi - \frac{1}{a^2} \nabla^2 \phi + 3 H_c \dot\phi +
V''(\phi_c) \  \phi = 0,
\end{equation}
where $V''(\phi_c) \equiv \left. \frac{\partial^2 V(\varphi)}{\partial
\varphi^2} \right|_{\phi_c}$. Here, the semiclassical approach
$\varphi(\vec x,t) =
\phi_c(t) + \phi(\vec x,t)$ was taken into account to describe the
quantum fluctuations $\phi(\vec x,t)$, with $<0|\varphi|0>=\phi_c(t)$
and $<0|\phi|0>=0$.
Here, $|0>$ denotes the vacuum state.
Furthermore,
$H_c(\phi_c) = {\dot a \over a}$ is
the Hubble parameter and $a$ is the scale factor of the universe.
During inflation, the universe is accelerated $\ddot a > 0$ and the
inflation ends when $\ddot a \sim 0$. The equation (\ref{1})
describes the matter field fluctuations (up to nonlinear terms
in $\phi$), on a globally homogeneous
and isotropic background spacetime described by a Friedmann - Robertson -
Walker metric
\begin{equation}\label{2}
ds^2 = - dt^2 + a^2(t) d\vec x^2.
\end{equation}
The eq. (\ref{1}) can be simplified by means of the map $\phi =
e^{-3/2 \int H_c dt} \chi$
\begin{equation}\label{qf}
\ddot\chi - a^{-2}\left[k^2_0(t) - k^2\right] \chi =0,
\end{equation}
where
\begin{equation}
k^2_0(t) = a^2 \left[\frac{9}{4} H^2_c +
\frac{3}{2} \dot H_c - V''[\phi_c(t)]\right]
\end{equation}
is the time - dependent wavenumber that
separates the infrared and the
ultraviolet sectors, which describe the two relevants scales during
inflation.
Since $\dot k_0(t) >0$, during inflation new and new modes
enters in the infrared sector such that in this sector the
frequency $\omega_k$ holds
\begin{equation}\label{frec}
\omega^2_k(t) = - a^{-2}\left[k^2_0(t) - k^2\right] < 0.
\end{equation}
As in previous works\cite{PRD96,CMS} one can write the redefined
matter field fluctuations in the infrared sector by means a Fourier
expansion which selects the long wavelengths modes $\chi_k
=e^{i \vec k. \vec x} \xi_k(t)$
(for $k \ll k_0$)
\begin{equation}
\chi_{cg} = \frac{1}{(2\pi)^{3/2}}
{\large \int} d^3 k \  \theta(\epsilon k_0 - k) \left[
a_k \chi_k + a^{\dagger} \chi^*_k \right].
\end{equation}
Thus, the coarse - grained field $\chi_{cg}$ describes
the matter field fluctuations
on the infrared sector. Here, $\epsilon \ll 1$ is a
dimensionless parameter and $a^{\dagger}_k, a_k$ are the creation and
destruction operators with the algebra $[a^{\dagger}_k , a^{\dagger}_{k'}]
= [a_k , a_{k'}]=0$ and
$[a^{\dagger}_k , a_{k'}]= i \delta(k-k')$.
The quantum to classical transition of the matter field fluctuations
in the infrared sector is well known\cite{PRD96,CQG99,CQG1,CQG2}.
The field $\chi_{cg}$
can be considered
as classical when the time dependent
modes of this sector holds the condition $\left|
{{\rm Im}[\xi_k(t)] \over {\rm Re}[\xi_k(t)]}\right|_{IR} \ll 1$.

The problem of the increasing number of degrees of freedom
for the matter field fluctuations in the
infrared sector in the context of thermodynamics is not
well studied.
To make a thermodynamic description
for the fluctuations in
the infrared sector we can
introduce the partition function $Z(\beta)$
\begin{eqnarray}
Z(\beta) &\simeq &{\large \int}^{\epsilon k_0}_{0}
\frac{d^3 k}{(2\pi)^3} e^{-\beta \omega_k(t)} \nonumber \\
& = & {\large \int}^{\omega_{\epsilon k_0}}_{\omega_{k=0}}
d \omega_k \  \rho(\omega_k) \  e^{-\beta \omega_{k}},\label{8}
\end{eqnarray}
where $\beta^{-1}$ plays the role of the
background ``temperature'' and
$\omega_{k}$ is the relevant frequency given by eq. (\ref{frec}).
Furthermore, I denote the squared frequency with the cut off
wavenumber $\epsilon k_0$
by $\omega^2_{\epsilon k_0} = -a^{-2} \left[k^2_0(1-\epsilon^2)\right]$.
Here, $\epsilon$ is a dimensionless parameter given by $k/k_0 \ll 1$.
In the framework of stochastic inflation the background is well
represented by the ultraviolet sector, where $\omega^2_k >0$.
In the semiclassical limit the frequency $\omega_k$ plays the
role of the energy for each mode with wavenumber $k$.
We are interested in the infrared sector. In this sector the wavenumbers
are very small with respect to $k_0$ ($k \ll k_0$), and the
frequency $\omega_k$ is imaginary pure
($\omega_k = \pm {\rm i} |\omega_k|$).
As we will see later, the parameter $\beta$ is also imaginary pure
and thus the argument of the exponential in eq. (\ref{8}) remains
real.
The function $\rho(\omega_k)$ gives the density of states with
frequency $\omega_k$ on the infrared sector
\begin{equation}
\rho(\omega_k) = \frac{1}{(2 \pi)^3} \left|\frac{d^3k}{d\omega_k}\right|=
\frac{k^2}{2 \pi^2} \left|\frac{dk}{d\omega_k}\right|,
\end{equation}
where $\left|{d^3k \over d\omega_k}\right|$ is the Jacobian of the
transformation,
so that
\begin{equation}
\left|\frac{ d \omega_k}{d k}\right| = \frac{
\left[k^2_0+a^2 \omega^2_k\right]^{1/2}}{a^{2} \  |\omega_k|},
\end{equation}
where $|\omega_k| = \left[\omega_k\omega^*_k\right]^{1/2}$ and the
asterisk denotes the complex conjugate.
The density of states with frequency $\omega_k$ is given by
\begin{equation}\label{rho}
\rho(\omega_k) = \frac{1}{2\pi^2} \left[k^2_0 + a^2(t)
\omega^2_k\right]^{1/2}
|\omega_k| \  a^2(t).
\end{equation}
The thermodynamics for systems with exponentially
growth of density of states was first considered
by Hagedorn in the framework of the hadron mass spectrum
in bootstrap models\cite{h1,h2}.
Energy added to a system can go either into
increasing the energy of existing states or into creating
new states. In the case of the matter field fluctuations in the
infrared sector
new and new states are created from the ultraviolet sector.

The ``temperature'' and the heat capacity are given by
\begin{eqnarray}
\beta &= & \left.\left|\frac{\partial {\rm ln}[\rho]}{\partial \omega_k}\right|
\right|_{k=\epsilon k_0}, \label{a}\\
C_V & = & - \left.\beta^2 \left[\frac{\partial^2 {\rm ln}[\rho]}{\partial
\omega^2_k}\right]^{-1}\right|_{k=\epsilon k_0}. \label{b}
\end{eqnarray}
The condition that the density of states rises superexponentially
is precisely that the second derivative in eq. (\ref{b}) be positive,
and $C_V$ thus be negative.
Systems with negative heat capacities
are thermodynamically unstable. They are placed in contact
with a heat bath and
will experience runaway heating or cooling.
If the density of states grows exponentially, an
inflow of energy at the Hagedorn temperature goes entirely
into producing new states, leaving the temperature constant.
If the density
of states grows superexponentially, the process is similar, but
the production of new states is so copious that an inflow of energy
actually drives the temperature down.
To simplify the notation, in the following I will
denote $\omega_{\epsilon k_0}$ as $\omega$.
For inflationary models one obtains
in the infrared sector
\begin{equation}
C_V  = \frac{-\mu^4 \left(\omega^2 \mu^2 + \mu^4 + 2 \omega^4\right)}{
\omega^4 \left(\mu^2+\omega^2\right)^4},
\end{equation}
where $\mu^2=k^2_0/a^2$. Furthermore, the inverse
of the effective temperature is
\begin{equation}
\beta \simeq \mp {\rm i} \frac{ \mu^2}{|\omega| \left(\mu^2+\omega^2\right)}.
\end{equation}
In the case we are studying, the ``thermal bath'' is described
by the ultraviolet sector, but it is not trully thermalized.
In the framework of supercooled inflation, $\beta^{-1}$ it is
not a trully temperature. Thus, it is imaginary pure.
Furthermore, the parameter $\beta$ describes the environment
of the infrared sector, here characterized by the ultraviolet
sector. The quantum nature of the matter field fluctuations in
the ultraviolet sector is another motivation for the parameter
$\beta$ to be imaginary pure.

Furthermore, the heat capacity gives information about the evolution
of the infrared sector, which is an unstable sector. If $C_V >0$, the
system distributes its energy in the existent states. The inverse
situation describes a system which increments very rapidly the number
of states.

{\em Thermodynamics for a de Sitter expansion}: As a firsth example
I will consider a scale factor $a \sim e^{H_0 t}$, where $H_0$ is the
Hubble parameter. For a de Sitter expansion this parameter is constant.
In a de Sitter expansion $\mu^2 = \nu^2 H^2_0$, where $\nu^2 = {9\over 4}-
{m^2 \over H^2_0}$.
The density of states is given by
\begin{equation}
\rho(\omega_k) \simeq \frac{1}{2\pi^2} |\omega_k| \left[
\nu^2 H^2_0 + \omega^2_k\right]^{1/2} \  e^{3 H_0 t}.
\end{equation}
From eq. (\ref{a}) one obtains the
inverse of the effective temperature for a de Sitter expansion
\begin{equation}
\beta \simeq \mp {\rm i} \frac{1}{\nu H_0 \epsilon^2 \sqrt{1-\epsilon^2}}
\simeq \mp \frac{\rm i}{\nu H_0 \epsilon^2}.
\end{equation}
which is imaginary pure and does not depends on time.
Furthermore, the heat capacity is [see eq. (\ref{b})]
\begin{equation}
C_V  \simeq  \frac{- \left(2+2 \epsilon^4 - 3 \epsilon^2\right)}{
\left(\epsilon^2 \nu H_0\right)^4 \left(1-\epsilon^2\right)^2}
\simeq -2  \left(\beta \beta^*\right)^2.
\end{equation}
Note that the heat capacity is negative but constant.
This means that, as we put energy into the infrared sector,
a greater and greater proportion of it is employed in the exponential
production of new states rather than
in increasing the energy of already existing states.

{\em Thermodynamics for a power - law expansion}: Now we consider
the case where the scale factor evolves as $a \sim t^p$.
In this case the Hubble parameter is $H_c= p/t$ and the
effective squared mass parameter becomes \cite{PRD96}
\begin{equation}
\mu^2(t) =t^{-2} \left[\frac{9}{4} p^2 - \frac{15}{2} p +2 \right].
\end{equation}
The density of states $\rho(\omega_k)$ is
\begin{equation}
\rho(\omega_k) \simeq \frac{|\omega_k|}{2\pi^2}
\left[K^2 t^{-2} + \omega^2_k\right] t^{3p},
\end{equation}
where $K = \sqrt{{9 \over 4} p^2 - {15 \over 3} p + 2}$.
Inflation holds when $K >0$, i.e., for $p > 3.04$.
Hence, the inverse of the effective temperature of the infrared sector is
[see eq. (\ref{a})]
\begin{equation}
\beta \simeq  \mp {\rm i} \frac{t}{K \epsilon^2 \sqrt{1-\epsilon^2}}
\simeq \mp \frac{{\rm i} t}{K \epsilon^2}.
\end{equation}
Furthermore, the heat capacity is obtained from eq. (\ref{b})
\begin{equation}
C_V = \frac{- \left(2+2\epsilon^4-3 \epsilon^2\right) t^4}{
\left(K \epsilon^2\right)^4 \left(1-\epsilon^2\right)^2}
\simeq -2 \left(\beta \beta^*\right)^2.\label{cv}
\end{equation}
The expression (\ref{cv}) for $C_V$
becomes more and more negative
with time,
due to the unstability of the infrared sector during inflation.
As was demonstrated in a previous
work\cite{PRD96}, in a power - law expansion
for the universe
the inflaton potential suppresses the dispersion of the quantum
fluctuations in a power - law expansion for the universe.
This result coincides with the quantum field prediction
and could be responsible for the very rapidly decreasing
of the heat capacity $C_V$.

{\em General comments}: If $(\beta\beta^*)^{-1/2}$ is the zero
mode temperature (or background temperature), the squared infrared
matter field fluctuations when the horizon entry will be
\begin{equation}
\left<\phi^2_{cg}\right>_{IR} \left.\simeq \frac{\epsilon^6}{
6\pi^2 (\beta\beta^*)^{3/2}} \left[\xi_{k=0}(t)\right]^2\right|_{t=t_*},
\end{equation}
where $t_*$ is the time when the horizon entry and $\xi_{k=0}(t)$ is
the solution for the zero mode equation of motion
$ \ddot\xi_0 - {\epsilon^2 \over \beta\beta^*} \xi_0 =0 $, so that
the amplitude for primordial power density perturbations should be a
function of the background temperature. In other words, if
${\cal P}_{\phi_{cg}}(t_*)=|\delta_k|^2$ is the power spectrum
for the matter field fluctuations, such that $\left<\phi^2_{cg}\right>_{IR}=
\int^{\epsilon k_0}_{0} {dk\over k} {\cal P}_{\phi_{cg}}$, hence
the density perturbations can be written as
$|\delta_k| = A(t_*) \  k^n$ ($n=3/2$), which agree with the best-fit
slope of COBE data: $n=1.2 \pm 0.3$. A best approximation could be
obtained from the exact solution for the equation for the modes:
$\ddot\xi_k + \left[k^2/a^2 - \epsilon^2/(\beta\beta^*)\right]\xi_k=0$.

To summarize, note that in the cases here developed ---
de Sitter and power - law expanding
universes ---
the heat capacity is negative. This is
because the density
of states in the infrared sector
grows exponentially or superexponentially
during inflation. Hence,
the increasing rate of states in the infrared sector is
more copious than the inflow of energy in this sector.
Rather an
increasing energy density, an increasing of $|C_V|$ (for
$C_V < 0$), gives a superproduction of the number of
degrees of freedom in the infrared sector.
The interesting fact is that in the both cases here studied one
obtains $\mu^2 (\beta\beta^*) \simeq \epsilon^2 $
and $C_V \simeq - 2(\beta\beta^*)^2$.
The main difference founded in the examples here studied is that
the heat capacity in the power - law expansion
model decreases very rapidly.
Of course, the mechanism for this suppression must be
understood from the thermodynamic analogy due to
systems with imaginary
temperatures and those with negative heat capacities occur
in different contexts\cite{c,d,e,f}, but their thermodynamic behavior has a
common physical basis.
The imaginary nature of $\beta$ can be a consequence of the
quantum nature of the matter field fluctuations in the infrared's
environment (i.e., of the ultraviolet sector).
The temporal dependence for $\beta$ and $C_V$ in the power-law
expansion
is due to the interaction of the inflaton field,
which manifests itself in the equation
of motion (\ref{qf}) for the quantum fluctuations
through the time-dependent mass parameter $\mu (t)\equiv k_0/a \sim t^{-1}$.
This effect generates --- for a power - law expanding universe ---
a superexponential increasing for the number of
degrees of freedom (i.e., the number of states) in the infrared sector,
rather a exponential increasing of states founded in a de Sitter expanding
universe due to $\beta$ and $C_V$ remain constant.
. So, in a power-law expanding universe the production of
new states is so copious that an inflow of energy actually drives
the background temperature $(\beta\beta^*)^{-1/2}$ down asymptotically to
zero, meanwhile in a de Sitter expansion the inflow of energy go entirely
into producing new states, leaving the background temperature constant.

Finally, super Hubble matter field fluctuations with negative heat
capacity during inflation describes exponential or superexponential growth
of the number of states, which is a characteristic of nonequilibrium
thermodynamical systems.  It shows that a more dynamical approach to
the statistical mechanics of inflation might be necessary.


\begin{thebibliography}{99}
\bibitem{lin} A.D.Linde, {\sl Particle Physics and Inflationary
Cosmology} (Harwood, Chur, Switzerland, 1990) and references therein.
\bibitem{star} A.A.Starobinsky, in {\sl Current Topics in
Field Theory, Quantum Gravity, and Strings}, ed. by H.J. de Vega and
N.S\'anchez, Lecture Notes in Physics {\ 226} (Springer, New York,
1986).
\bibitem{goncharov}  A.S.Goncharov and A.D.Linde, Sov.J.Part.Nucl {\bf 17}
369 (1986).
\bibitem{c} S. Carlip, Phys. Rev. Lett. {\bf 79}, 4071 (1997).
\bibitem{p} T. Padmanabhan, Phys. Rev. Lett. {\bf 81},
4297 (1998).
\bibitem{Hab} S.Habib, Phys. Rev. {\bf D46}, 2408 (1992).
\bibitem{PRD96} M. Bellini, H. Casini, R. Montemayor, P. Sisterna,
Phys. Rev. {\bf D54}, 7172 (1996).
\bibitem{CMS} M. Bellini, Phys. Rev.{\bf D61}, 107301 (2000).
\bibitem{CQG99} M. Bellini, Class. Quantum Grav. {\bf 16}, 2393 (1999);
M. Bellini, Nucl. Phys. {\bf B563}, 245 (1999).
\bibitem{CQG1} D. Polarski and A. A. Starobinsky, Class. Quant. Grav.
{\bf 13}, 377 (1996).
\bibitem{CQG2} C. Kiefer, J. Lesgourdes, D. Polarski, and
A. A. Starobinsky, Class. Quant. Grav. {\bf 15}, L67 (1998).
\bibitem{h1} R. Hagedorn, Nuovo Cimento Suppl. {\bf 3}, 147 (1965).
\bibitem{h2} R. Hagedorn, Nuovo Cimento A {\bf 56}, 1027 (1968).
\bibitem{d} P. Hertel and W. Thirring, Ann. Phys. (N.Y.) {\bf 63},
520 (1971).
\bibitem{e} D. Lynden-Bell and R. M. Lynden-Bell, Mon. Not. R. Astron.
Soc. {\bf 181}, 405 (1977).
\bibitem{f} P. T. Landsberg and R. P. Woodard, J. Stat. Phys. {\bf 72},
361 (1993).
\end{thebibliography}
\end{document}